\documentclass[3p]
{elsarticle}
\newcommand{\be}{\begin{equation}}
\newcommand{\ee}{\end{equation}}
\newcommand{\beq}{\begin{eqnarray}}
\newcommand{\eeq}{\end{eqnarray}}
\newcommand{\lb}[1]{\label{#1}}
\newcommand{\sty}{\scriptstyle}

\newcommand{\e}{{\mathrm{e}}}
\newcommand{\dd}{{\mathrm{d}}}
\usepackage{amsmath}
\usepackage{txfonts}
\usepackage{epsfig}
\usepackage[obeyspaces,hyphens]{url}
\usepackage{graphicx}
\begin{document}
\title{Testing the Goodwin growth-cycle macroeconomic dynamics in Brazil}

\author[ibge]{N.J.\ Moura Jr.}
\ead{newton.junior@ibge.gov.br}

\author[if]{Marcelo B.\ Ribeiro\corref{cor1}}
\ead{mbr@if.ufrj.br}

\cortext[cor1]{Corresponding author}
\address[ibge]{Instituto Brasileiro de Geografia e
               Estat\'{\i}stica -- IBGE, Rio de Janeiro, Brazil}
\address[if]{Instituto de F\'{\i}sica, Universidade Federal do
             Rio de Janeiro -- UFRJ, Rio de Janeiro, Brazil}

\begin{abstract}
This paper discusses the empirical validity of
Goodwin's (1967) macroeconomic model of growth with
cycles by assuming that the individual income
distribution of the Brazilian society is described by
the Gompertz-Pareto distribution (GPD). This is formed
by the combination of the Gompertz curve, representing
the overwhelming majority of the population ($\sim
\,$99\%), with the Pareto power law, representing the
tiny richest part ($\sim \,$1\%). In line with Goodwin's
original model, we identify the Gompertzian part with
the workers and the Paretian component with the class
of capitalists. Since the GPD parameters are obtained
for each year and the Goodwin macroeconomics is a time
evolving model, we use previously determined, and further
extended here, Brazilian GPD parameters, as well as
unemployment data, to study the time evolution of these
quantities in Brazil from 1981 to 2009 by means of the
Goodwin dynamics. This is done in the original Goodwin
model and an extension advanced by Desai et al.\ (2006).
As far as Brazilian data is concerned, our results show
partial qualitative and quantitative agreement with both
models in the studied time period, although the original
one provides better data fit. Nevertheless, both models
fall short of a good empirical agreement as they predict
single center cycles which were not found in the data. We
discuss the specific points where the Goodwin dynamics
must be improved in order to provide a more realistic
representation of the dynamics of economic systems.
\end{abstract}

\maketitle
\biboptions{sort&compress}
\section{Introduction}\lb{intro}

It has been noted long ago by Karl Marx that capitalist production
grows on cycles of booms and busts. During a boom, profits increase
and unemployment decreases since the workers are able to get better
jobs and higher salaries due to shortage of manpower to feed the
growing production. However, this boom is followed by a bust since
less unemployment reduces the profit margin, whose recovery is
achieved by a higher unemployment and a reduction of the workers'
bargaining power. Smaller salaries increase the profit margin
leading to renewed investment and then a new boom starts, being
followed by another bust, and so on \cite[Chap.\ 25, Sect.\ 1]
{marx}. 

A century later Richard Goodwin \cite{g67} proposed a
mathematical model which attempts to capture the essence of Marx's
dynamics described above. In this model the basic dynamics of a
capitalist society, as qualitatively described by Marx, is modeled
by means of a modified Lotka-Volterra model where predator and prey
are represented by workers and capitalists. Goodwin replaced the
classic Lotka-Volterra dynamics of number of predators and preys by
two new variables $u$ and $v$, the former giving the workers' share
of total production, which is an indirect way of describing the
profit margin of capitalists, and $v$ representing the employment
rate, which is an indirect way of describing the share of those
marginalized by the production, the unemployed workers, that is,
the industrial reserve army of labor in Marx's terminology.
In a boom the employment rate $v$ increases and the workers' share
$u$ starts to increase after a time lag, meaning a decrease in profit
margin. When employment rate is at its maximum this corresponds to
the lowest profit margin, then the burst phase starts with a
decrease in $v$. At this point $u$ had already started diminishing.
The essence of the model is captured as a closed orbit in the
$u$-$v$ phase space. Clearly these two variables are out of phase
in time \cite[pp.\ 458-464]{gan97}.

Although the brief description given above appears to indicate
that Goodwin was able to capture Marx's observations, the model
has in fact several shortcomings, the most severe one being its
inability to predict quantitatively the above described dynamics
(see below). The model was presented simply as an heuristic
reasoning capable of giving a mathematical dressing to Marx's
ideas. It was born out as a vision of the world rather than from
a real-world data-inspired model in a physical sense. Despite
this, or, perhaps, because of this, since its formulation Goodwin's
model has attracted considerable theoretical attention in some
economic circles and several variations of the original model
were proposed \cite[see][and references therein]{g84,sp,sordi,
sordi01,vene,dhmp,va,vm,dibeh,kv,cst,tc,ac,ba,keen95}.

However, interestingly enough, almost half a century after its
proposal, attempts to actually \textit{test} this model
empirically are still extremely limited. Although Goodwin's
growth-cycle model is certainly influential in view of the
number of \textit{theoretical} follow-up papers cited
above, studies seeking to establish its empirical soundness are
limited only to Refs.\ \cite{at,desai,solow,harvie,va,moreno,
simon,mome}. This is a surprisingly short list when we consider
the time span since the model's initial proposal. So little
interest in empirically checking the model, especially among
those who appear to have been seduced by its conceptual aspects,
is even more surprising if we bear in mind that for the last
30 years or so we have been living in an era where large economic
databases are easily available digitally, so large-scale checking
of this model against empirical data ceased long ago to pose an
insurmountable barrier. Besides, even the very few studies
which actually attempted that, all point to severe empirical
limitations of the model, ranging from partial qualitative
acceptance to total quantitative rejection. From an econophysics
viewpoint it is curious that a model with such a poor empirical
record became so influential.

Despite this, the model does have some general empirical
correspondence to reality on a qualitative level and this
justifies further empirical studies with different databases,
data handling methods and/or data type approaches. The basic
aim must lie in identifying as clearly as possible where the
model performs poorly in order to propose amendments and
modifications. Any model, especially those theoretically seducing,
can only remain of interest if it passes the test of experience,
if it survives confronting its predictions with empirical data.
If it does not survive this test the model must be modified,
or abandoned.

This paper seeks to perform an empirical study of the Goodwin
growth-cycle model using individual income data of Brazil. The
study presented here was directly motivated by our previous
experience in modeling Brazil's income distribution, whose
results suggested a Goodwin type oscillation in the share of
the two income classes detected in the data \cite{nm09,fnm10}.
Building upon our previous experience with this database, we
obtained yearly values of the two main variables of the
Goodwin model, the {labor share} $u$ and the
{employment rate} $v$. Nevertheless, differently from
all previous approaches for testing Goodwin's model, here the
labor share was obtained by modeling the individual income
distribution data with the Gompertz-Pareto distribution (GPD) and
identifying $u$ with the Gompertzian, less wealthy, part of the
distribution \cite{fnm10}. The employment rate was also estimated
from the same database, that is, from Brazil's income distribution,
using the concept of \textit{effective unemployment}.

We show that from 1981 to 2009 $u$ and $v$ do cycle in a form
bearing similarities to what the Goodwin model predicts, that is,
closed cycles. However, our results show the absence of a single
cycling center and also are in complete disagreement with the ones
for Brazil as reported by Ref.\ \cite{mome}, whose analysis employed
Harvie's method \cite{harvie}. In addition, we attempted to see if
our findings bring empirical support to the Desai-Henry-Mosley-Pemberton
(DHMP) extension of the original model \cite{dhmp}. Our results show
that this particular variation of the Goodwin dynamics has some
empirical soundness, although it provides a somewhat poorer data fit
as compared to the original model and also leaves three parameters 
to be determined by other, still unknown, means than the ones studied
here, whereas the original model leaves two parameters in a similar
situation. We conclude that these two models provide partial qualitative
and quantitative agreement with real data, at least as far as empirical
data from Brazil are concerned, but both of them, and perhaps all
variations of the original Goodwin growth-cycle dynamics, require
important modifications and amendments before they can be considered
viable representations of the real dynamics of economic systems. 

The paper is organized as follows. Sect.\ \ref{model} presents
a brief review of the original Goodwin model and its DHMP extension,
focusing mostly on their dynamical equations, although some discussion
about the underlying economic hypotheses and foundations of the original
model is also presented. In Sect.\ \ref{gpd}, after a short
discussion about methodology, we review the main equations behind the
GPD. Sect.\ \ref{data-orbits} analyzes the individual income data of
Brazil and presents the $u$-$v$ orbits in the 1981-2009 time period.
Sect.\ \ref{5} provides time variations of the employment rate as
compared to workers' share so that line fittings allow us to determine
some of the unknown parameters of both models. Finally, Sect.\ \ref{fim}
discusses the results and presents our conclusions.

\section{The Goodwin growth-cycle macro-economic dynamics}\lb{model}

\subsection{The original growth-cycle model}\lb{growth}

The model proposed by Goodwin is essentially a Lotka-Volterra
predator-prey system of first order ordinary differential
equations which can be written as follows \cite{g67,harvie,dhmp},
\be
\dot{u} = \left[ - \left( a + d \right)
          + h v \right] u, 
\lb{u1}
\ee
\be
\dot{v} = \left[ \displaystyle \frac{100-u}{c}
          - \left( a + b \right) \right] v,
\lb{v1}
\ee
where the dot denotes the time differentiation $\mathrm{d}/\mathrm{d}t$.
The five constants $a, b, c, d, h$ come from the economic hypotheses
of the model and are supposed to obey the following conditions
\cite{gan97,harvie},
\begin{equation}
 \left\{ \begin{array}{ll}
              c>0, \\
	      h>0, \\
	      (a+d)>0, \\
	      (a+b) \, c<100.
         \end{array}
  \right.
\label{pars}
\end{equation}
The solution of equations (\ref{u1}) and (\ref{v1}) produces a family of
closed orbits with period $T$, all having the point $(u_c, v_c)$ as
their unique center, according to the following equations \cite{harvie},
\begin{equation}
 \left\{ \begin{array}{ll}
              u_c=100- (a+b)c, \\
	      v_c=(a+d)/h, \\
	      T= {2 \pi} \, \big. \big/ {\sqrt{(a+d)[100/c-(a+b)]}}.
         \end{array}
  \right.
\label{center}
\end{equation}
Since $u$ is the \textit{percentage share of labor, or workers, in
national income} and $v$ represents the \textit{proportion of labor
force employed}, they both should lie in the [0, 100] interval. Here
we follow the normalization adopted in Refs.\ \cite{nm09,fnm10} and
shall refer to the maximum share, or proportion, as 100\%. The upper
singular point $v_s$ for the employment proportion is reached when
$\dot{v}=0$, then $u_s=100-c(a+b)$. Similarly, when $\dot{u}=0$ we
have $v_s=(a+d)/h$. However, if $(a+b)$ is negative, then $u_s>100$,
which, in principle, should not be allowed (for a conceptually
possible, but so far untested, exception, see Ref.\ \cite{gan97},
p.\ 461). Similarly, it is possible to have $v_s>100$.

In this model $u$ represents the population density of predators
whereas $v$ represents the prey population density. This can be
seen as follows. When $u=0$, $\dot{u}=0$ and $\dot{v}>0$. In other
words, $u$ remains equal to zero whereas $v$ grows without bound,
a situation happening to the prey population $v$ in the absence of
predators $u$. On the other hand, when $v=0$, equations (\ref{u1}) and
(\ref{v1}) together with conditions (\ref{pars}) show that
$\dot{v}=0$ and $\dot{u}<0$. So, without prey ($v=0$), the
predator population decreases ($\dot{u}<0$).

The model is defined in terms of five parameters. However, once they
are grouped as below,
\be
\left\{ \begin{array}{ll}
a_1=(a+d), \\
a_2=(100/c)-(a+b), \\
b_1=h, \\
b_2=100/c,
        \end{array}
\right.
\lb{ctes}
\ee
they allow equations (\ref{u1}) and (\ref{v1}) to be rewritten in the form
of the classical Lotka-Volterra equations
\cite{gan97},
\begin{equation}
  {\dot{u}}/{u} = - a_1 + b_1 \, v,
\lb{lv1}
\ee
\be
  {\dot{v}}/{v} = a_2 - b_2 \, u, 
\label{lv2}
\end{equation}
that is, in terms of four parameters which could, in principle, be
determined observationally, provided that both variables and their
derivatives are obtained from real data.

\subsection{The Desai-Henry-Mosley-Pemberton (DHMP) extension}\lb{dhmp}

Desai et al.\ \cite{dhmp} noted that the original Goodwin model
can produce solutions outside the $u$-$v$ domain $[0,100] \times
[0,100]$ because, as seen above, both $u_s$ and $v_s$ can grow above
100. This is the main reason which led them to propose a modified
version of Goodwin's original model, dubbed here as the DHMP
extension. They also relaxed two other economic hypotheses assumed
in the original model. So, in the DHMP extension all profits are not
always invested and the Phillips curve, relating unemployment and
inflation rate, is non-linear. Thus, the final equations yield,
\be
  \dot{u}=\left[ -( \bar{a}+ \bar{d})+ \bar{h}{(100-v)}^{\delta}
  \right] u,
  \lb{u-desai}
\ee
\be
 \dot{v}=\left\{ \left[ -\lambda \ln (100-\bar{u})-(\bar{a}+\bar{b})
 \right] + \lambda \ln (\bar{u} -u) \right\} v,
  \lb{v-desai}
\ee
where $\bar{a}$, $\bar{b}$, $\bar{d}$, $\bar{h}$, $\delta$,
$\lambda$, $\bar{u}$ are constants obeying the following constraints,
\begin{equation}
 \left\{ \begin{array}{ll}
          \delta>0, \\
          \lambda>0, \\
          u<\bar{u}<100, \\
          \bar{h}< (\bar{a}+\bar{d}), \\
          (\bar{a}+\bar{b})< \lambda \ln \left( \displaystyle
            \frac{\bar{u}}{100- \bar{u}} \right), \\
   \displaystyle \left( \frac{\bar{u}}{100- \bar{u}} \right)>1.
         \end{array}
  \right.
\label{pars2}
\end{equation}
Ref.\ \cite{dhmp} gives a clear meaning to the parameter
$\bar{u}$ as being ``the maximum share of labor that capitalists
would tolerate'', ``typically'' given by the last constraint equation
in the set of expressions (\ref{pars2}) above. Clearly this implies
that $\bar{u}>50$\%. One must also note that both the original
Goodwin model and its DHMP extension consider that the labor share
and profits are not given in terms of money, but in real terms. As
we shall see below, this requirement does not pose a problem for our
approach since our variables are currency independent \cite[see][]{nm09}.

As seen above, the DHMP extension of Goodwin's growth cycle model
is defined by seven parameters which can be grouped as below,
\be
\left\{ \begin{array}{ll}
\bar{a}_1=(\bar{a}+\bar{d}), \\
\bar{a}_2= \lambda \ln (100-\bar{u})+(\bar{a}+\bar{b}), \\
\bar{b}_1=\bar{h}, \\
\bar{b}_2=\lambda,
        \end{array}
\right.
\lb{ctes2}
\ee
allowing us to rewrite equations (\ref{u-desai}) and (\ref{v-desai})
as follows,
\begin{equation}
  {\dot{u}}/{u} = - \bar{a}_1 + \bar{b}_1 \, V^\delta,
\lb{lv-d1}
\ee
\be
  {\dot{v}}/{v} = -\bar{a}_2 + \bar{b}_2 \, \ln \mathcal{U}, 
\label{lv-d2}
\end{equation}
where
\be
 \mathcal{U} \equiv \bar{u} - u,
\lb{U}
\ee
and the \textit{unemployment rate} given by,
\be
 V \equiv 100 - v.
\lb{V}
\ee

Although the basic motivation for the DHMP extension was to avoid
the variables of the model having values above 100\%, this difficulty
can be avoided if both $u$ and $v$ are defined by real data, in which
case the desired threshold will be achieved by construction. Besides,
the DHMP model has the additional disadvantage of requiring seven,
rather than five, unknown parameters. 

\subsection{Interpretation of the conflicting variables}\lb{conflict}

As seen above the Goodwin model is essentially a predator - prey
type one and this means that its two variables represent the
opposing, but interdependent, nature of a predator - prey conflict.
This is the reason why this model is also known as ``Goodwin's class
struggle model.'' The nature of this ``struggle'' arises from the
possible ways we interpret its variables.

On one hand, the employment rate $v$ can be identified with
the workers' class and the profit share of the ``capitalists''
is then given by,
\be
U \equiv 100-u.
\lb{inv}
\ee
In other words,
$U$ is the share of total national income obtained
by the class that controls the capital, the investors. In this
case the conflict is between the workers and the investors
(capitalists). That can be seen in the light of a change of
variables such that when $u=0$, $\dot{u}=0$, $\dot{{U}}=0$
and $\dot{v}>0$, meaning that when the profit ${U}$
attained by investors remains constant, i.e., $\dot{U}=0$,
the workers' share $v$ grows without bound and represents the
prey, whereas the investors ${U}$ are in the role of
predators. Here ${U}$ is assumed to have a maximum value
equal to 100\%.

On the other hand, following Solow \cite{solow}, employed workers
can be identified with the workers' share $u$ and unemployed workers
with the variable $V$. In this case the conflict is between employed
and unemployed workers. When $u=0$, $\dot{u}=0$ and $\dot{V}<0$.
This is consistent with the employed workforce $u$ in the role of
prey, the unemployed workers $V$ being identified with the predators
and the investors as passive non-players.

However, these interpretations should not be taken at their face
values as they are dependent on the conditions given by equations
(\ref{pars}). Such parameter constraints were, however, not
established from an analysis of real-world data, but came from
entirely heuristic, and so far very poorly tested, reasoning.
In addition, since as seen above the variables $u$ and $v$ can be
identified in more than one way, this means that such
interpretations must be done with care and always in the light of
real-world data analysis and not on a speculative basis. As further
emphasis of these difficulties, one may even argue that the
constants of the model may not be constants at all, but time
dependent variables themselves (see below).

\subsection{Origins of the Goodwin model}\lb{origin}

As noted above, when developing his model Goodwin aimed at putting
in mathematical form Marx's conceptual ideas about cycles in
capitalism. However, as pointed out by Keen \cite{keen2012}, Goodwin
also wished to show how cyclical behavior could arise from very
simple economic hypotheses. Next we shall present a simple derivation
of the model in order to highlight that it results from an extremely
simplified representation of the economy.

Let $K$ be the amount of \textit{fixed capital} (plant and equipment)
and $Y$ the \textit{output} that an economy can generate. The
\textit{output to capital ratio} $\sigma$ clearly varies over time in
a country, but let us consider it a constant as a first approximation
and write it as follows,
\be \sigma={K}/{Y}. \lb{o1} \ee
If $L$ is the \textit{amount of labor} for a given output, one can also
assume as first approximation a constant \textit{output to labor ratio}
$a$, that is, 
\be a={Y}/{L}. \lb{o2} \ee
The amount of labor can be written in terms of the \textit{population}
$N$ and the employment rate $v$ as follows,
\be L=Nv. \lb{o3} \ee
Let $w$ be the \textit{average wage value}. Then the \textit{wage bill},
that is, the total amount of wages in an economy is given by,
\be W=wL. \lb{o4} \ee
At a first approximation the employment rate can be related to
the rise of wages as follows,
\be {\dot{w}}/{w}=f_1(v). \lb{o6} \ee
Since the wage share $u$ is given by,
\be u={wL}/{Y}, \lb{o7} \ee
remembering that $a$ is constant, equation (\ref{o6}) becomes,
\be {\dot{u}}/{u} = f_1(v). \lb{o8} \ee
This expression reduces to equation (\ref{lv1}) if $f_1(v)$ is assumed
to be a linear function.

The \textit{profit level} $P$ is given by,
\be P=Y-W.  \lb{o5} \ee
As a first approximation all profits are invested, so the
\textit{profit share} $P/Y$ is the \textit{investment} $\Upsilon$.
Hence,
\be \Upsilon ={P}/{Y}=100-({W}/{Y})=100-u=U. \lb{o9} \ee
Here the unit was changed to 100\% due to our previous choice of
normalization. The \textit{profit rate} $\pi$ is given by,
\be \pi={P}/{K}, \; \; \; \Longrightarrow \; \; \;
  \Upsilon=\sigma \pi =100-u, \lb{o10}
\ee
which can be rewritten in functional form as below,
\be \Upsilon=f_2(u). \lb{o11} \ee
Investment is also the \textit{rate of change of capital}
$\dot{K}/K$. So,
\be \Upsilon=\frac{\dot{K}}{K}=\frac{\dot{Y}}{Y}=\frac{\dot{v}}{v}+
    \mathrm{const.} \, , \lb{o12}
\ee
where the constant comes from the hypothesis of a steady labor supply,
e.g., $L$ changes exponentially. Summing up we have that,
\be {\dot{v}}/{v}=f_2(u), \lb{o13} \ee
which reduces to equation (\ref{lv2}) if $f_2(u)$ is assumed linear.    

Clearly the model results from extremely simple specifications
of the economy. But, it is so simple that it cannot reproduce the
frequency properties of output growth in a certain time period or
the distribution of recession sizes and duration. However, the
dynamic stochastic general equilibrium (DSGE) models of cycles
adopted by current neoclassical economics cannot do so either
\cite{solow2,ft,miro}, hence what is remarkable is that the very
restricted model proposed by Goodwin finds any empirical support in
real data \cite{keen2012}.

\section{The Gompertz-Pareto income distribution}\lb{gpd}

Econophysics is a new research field whose problems interest both
economists and physicists. However, when physicists approach a
problem traditionally dealt with by economists, they do so under
a very different modeling perspective. Although it is uncommon to
find methodological issues discussed in physics papers, considering
the hybrid nature of econophysics and the theoretical crisis of the
current mainstream economic thought \cite{solow2,ft,miro,kirman,keen,
keen11,c10,wea,krug,timeb,harvey,soos,bubble,hb}, it is worthwhile to
emphasize the differences in methodological perspectives between
physics and economics regarding model building and, especially, model
abandoning. We have already expressed some of our thoughts on this
topic in Ref.\ \cite[Sect.\ 3]{nm09}, but a few more words are worth
saying before we review our approach to the income distribution problem.

Econophysics was born and remains a branch of physics
\cite{pt2005,s10,s10p}, employing, therefore, its centuries old
proven epistemological methodology. It considers a scientific
theory as being made by laws of nature, which are theoretical
constructs, often expressed in mathematical language, that capture
regularities, processes, structures and interrelationships of
reality. Successful physical laws provide good empirical
\textit{representations}, or images, of the real world, of nature,
and allow us to reach predictions regarding the outcomes of
processes that do go on in nature. However, by being images of
nature, these laws are obviously limited and, hence, they will
always provide imperfect representations. The only way we can
ascertain how imperfect they are is by practice, i.e., by
creating pragmatic measures of the adequacies of these laws,
always empirically comparing their predictions with what occurs
in the real world \cite{rv98}. In other words, good laws provide
good predictions, bad laws provide bad predictions. This has nothing
to do with the extensive use of mathematics by physical theories.
Mathematics is a language, a tool of formal logic, and by itself
has no a priori relationship with physical, or social, reality.
Physicists \textit{choose} if and which mathematical tools are
required to express something observed in nature.\footnote{Here
we take a viewpoint different from Lawson's \cite{lawson}
regarding the role of mathematics in economics, a viewpoint based
on the larger experience of other sciences which successfully
adopted mathematical modeling, especially, but not restricted, to
physics. The obvious failures of mathematical modeling in economics
is a problem specific to academic economics because it misinterpreted
the role of theoretical thinking by means of a continuing excessive
emphasis in theoretical introspection parallel to a strong
downplaying of the empirical certification of models. Hudson
\cite{hudson} provides an interesting account of why and how
academic economics reached this present state of affairs. One
must note that the impressive achievements of the 20th century in
theoretical physics would never had occurred if physicists had
ignored empirics to the extent that academic economists do.
}

Since our understanding of the theories behind these laws changes
with time, the same occurring with the measures of adequacies due
to technological advances, we must keep measuring the adequacies
of these laws by perfecting old measures as well as creating new
ones, that is, constantly updating our theories and models through
practice in order to find their limits of validity. The theoretical
aspects behind these laws, even their metaphysical presuppositions,
must also be perfected by shedding the inappropriate elements so
that the appropriate residue remains, in a process very similar to
Darwin's natural selection. And, if there is no appropriate residue
left the theoretical construct is abandoned, becoming extinct
\cite{rv07}. Under this viewpoint, a model is a more restricted
theoretical construct, taking one or two elements above -- regularities,
processes, structures and interrelationships --, but not all of
them. Nevertheless, a model is also subject to measures of adequacy
and since they incorporate less elements than a theory, it suffers
a more rapid process of perfection by selection as well as extinction.

Physicists have been following this methodological approach for
centuries and as a consequence they have amassed a large number 
of physical theories that were perfected by generations of
physicists, who kept their appropriate kernels but changed
their original elements in various degrees, and also to many
other theories which are now superseded. Theoretical pluralism
is tacitly accepted as an essential element for the development of
physics. Real science starts from observation of nature, either
physical or social, and any theoretical discussion must keep
referring back to empirics, a factor that limits and guides any
theoretical debate, leading to healthy refining, replacing or
even abandoning of theories and models \cite{boltz}.

However, it seems that this methodological viewpoint regarding
model checking has not been adopted by a sizable number of
economists. Econophysicists are often perplexed to witness how
often economists confuse their models with reality, showing a
behavior which was already described as `scientific dogmatism'
\cite{rv98}. Thus, they would often disregard startling obvious
empirical facts rather than change or dismiss their inappropriate
theories or models \cite{b08,b09}, showing to a large extent an
absolute devotion to theoretical economic constructs, especially
an empirically unwarranted obsession with equilibrium, in parallel
to little or no empirical interest, often keeping such a
theoretical worship even when empirical evidence that might support
the theory is absent. Worse still, even when there is evidence
that directly contradicts what would be predicted to occur by
applying the theories \cite[pp.\ 2-5]{sinha}. Some would say 
this phenomenon is due to `ideological assumptions', disguised
visions of the world under scientific pretenses \cite{hudson}.
Others call this behavioral mode `cargo cult
economics' \cite{cargo1} in reference to the famous Feynman
speech about methodologically inadequate, or false science
\cite{feynman,georges}. Nevertheless, the epistemological ideas
above, adopted by physicists a long time ago, are apparently
being slowly absorbed into the economic thought \cite{birks,ness}.

Having stated our methodological viewpoints, next we shall review
the basic hypotheses and equations behind the GPD as advanced in
Refs.\ \cite{nm09,fnm10}.

\subsection{Definitions}\lb{basic}

Let $\mathcal{F}(x)$ be the {\it cumulative income distribution}
giving the probability that an individual receives an income less
than or equal to $x$. Then the \textit{complementary cumulative
income distribution} $F(x)$ gives the probability that an individual
receives an income equal to or greater than $x$. It then follows
that $\mathcal{F}(x)$ and $F(x)$ are related as follows, 
\be \mathcal{F}(x)+F(x)=100,
    \lb{ff}
    \ee
where the maximum probability is taken to be 100\%. Here $x$ is a
normalized income obtained by dividing the nominal income values
by some suitable nominal income average \cite{nm09}. If both
functions $\mathcal{F}(x)$ and $F(x)$ are continuous and have
continuous derivatives for all values of $x$, we have that,
\be \dd\mathcal{F}(x)/\dd x = f(x), \; \; \; \; \dd F(x)/\dd x=-f(x),
    \lb{c}
\ee
and
\be \int_0^\infty f(x):\dd x=100, \lb{norm1}
\ee
where $f(x)$ is the {\it probability density function of
individual income}. Thus, $f(x),\dd x$ is the fraction of
individuals with income between $x$ and $x+\dd x$. The equations
above lead to the following results,
\be \mathcal{F}(x) - \mathcal{F}(0) = \int_0^x f(w) \: \dd w,
\qquad
F(x) - F(\infty) = \int_x^\infty f(w) \: \dd w,
    \lb{4}
\ee
whose boundary conditions are,
\be \left\{ \begin{array}{lclcl}
    \mathcal{F}(0) & = & {F}(\infty) & \cong & 0, \\
    \mathcal{F}(\infty) & = & {F}(0) & \cong & 100.
    \end{array}
    \right.
   \lb{condi}
\ee
Clearly both $\mathcal{F}(x)$ and $F(x)$ vary from 0 to 100.

\subsection{The Gompertz-Pareto distribution (GPD)}\lb{gpdis}

The GPD was proposed in Ref.\ \cite{nm09} and discussed in detail
in Ref.\ \cite{fnm10}. Its complementary cumulative distribution
is formed by the combination of two functions which can be
identified with the two main classes forming most modern societies,
workers and investors (capitalists). The first component describes
the lower part of the distribution, that is, those who survive
solely on their wages, the workers, and is given by a
\textit{Gompertz curve}. The second component of the complementary
cumulative distribution describes the tail of the distribution by
means of the \textit{Pareto power law} and represents the investors,
that is, the rich capitalists. Then we have that,
\be F(x)= \left\{ \begin{array}{lcllc}
          G(x) & = & \e^{ \displaystyle \e^{(A-Bx)} },
	  & \; \; ( \: 0 \le x < x_{\sty t}), 
	  & \; \mbox{(Gompertz)} \\ \\
	  P(x) & = &
   {(x_{\sty t})}^{ \alpha} \;
          \e^{ \displaystyle \e^{(A-Bx_{\sty t})} }
	  \; x^{ - \alpha},
	  & \; \; (x_{\sty t} \le x \le \infty),
	  & \; \mbox{(Pareto)} \\
	  \end{array}
	  \right.
          \lb{distro}
\ee
and the cumulative income distribution may be written as
below,
\be \mathcal{F}(x)= \left\{
    \begin{array}{ll}
      \mathcal{G}(x)=100-\e^{ \displaystyle \e^{(A-Bx)} },
      & \; \; ( \: 0 \le x < x_{\sty t}), \\ \\
      \mathcal{P}(x)=100-
   {(x_{\sty t})}^{ \alpha} \;
          \e^{ \displaystyle \e^{(A-Bx_{\sty t})} }
      \; x^{ - \alpha},
      & \; \; (x_{\sty t} \le x \le \infty). \\
    \end{array}
    \right.
    \lb{disto1}
\ee
Here $x_{\sty t}$ is the income value threshold of the Pareto
region, $\alpha$ is the \textit{Pareto index} describing the
slope of the power law tail, $B$ is a third parameter characterizing
the slope of the Gompertz curve and $A$ is a number whose value is
set by boundary conditions, as follows. Since $G(x)=\exp \left[
\exp (A-Bx) \right]$, the condition (\ref{condi}) implies $G(0)=
100$, then we have that,
\be
   A=\ln \left( \ln 100 \right)=1.5272 \, .
   \lb{a}
\ee
The term ${(x_{\sty t})}^{ \alpha} \; \e^{ \displaystyle \e^{(A-
Bx_{\sty t})} }$ is the normalization constant of the Pareto
power law and comes as a consequence of condition (\ref{norm1}),
as well as the continuity of functions (\ref{distro}) across the
frontier between the Gompertz and Pareto regions, defined to be
$x=x_t$. 

The equations above allow us to find the expressions for the
probability density income distribution, 
\be f(x)= \left\{ \begin{array}{lclc}
          g(x) & = & B \; \e^{(A-Bx)} \; 
          \e^{ \displaystyle \e^{(A-Bx)} },
	  & \; \; ( \: 0 \le x < x_{\sty t}), \\  \\
          p(x) & = & \alpha \;
   {(x_{\sty t})}^{ \alpha} \;
          \e^{ \displaystyle \e^{(A-Bx_{\sty t})} }
	  \; x^{^{\scriptstyle
	  -(1+\alpha )}}, & \; \; (x_{\sty t} \le x \le
	  \infty),  \\
	  \end{array}
	  \right.
          \lb{distro2}
\ee
as well as the average income of the whole population described by
the GPD,
\be
  \langle x \rangle =\frac{1}{100}\int_0^\infty x \: f(x)
  \: \dd x = \frac{1}{100} \left[ \mathcal{I}(x_t) + \frac{\alpha \:
  x_{\sty t}}{(\alpha -1)} \e^{ \displaystyle \e^{(A-Bx_t)} } \right],
      \lb{avg}
\ee
where, 
\be \mathcal{I}(x)
    \equiv \int_0^x w \, g(w) \, \dd w=
    \int_0^x w \: B \: \e^{(A-Bw)} \;
    \e^{ \displaystyle \e^{(A-Bw)}} \dd w. 
    \lb{I}
\ee
The parameters $\alpha$, $x_{\sty t}$ and $B$ are all positive
and they fully characterize the GPD. However, due to convergence
requirements \cite{nm09}, the expression (\ref{avg}) for the
average income is only valid if $ \alpha > 1$. Both $\alpha$ and
$B$ can be determined by linear data fitting since equations
(\ref{distro}) can be linearized. However, $x_{\sty t}$ is
independently found under the constraint that the boundary
condition (\ref{a}) is satisfied to whatever degree of precision
the available data allow. 

The \textit{Lorenz curve} of the GPD has its X-axis given by
the cumulative income distribution function $\mathcal{F}(x)$,
whereas the first-moment distribution function $\mathcal{F}_1(x)$
defines its Y-axis. Accordingly, they can be written as follows
\cite{fnm10},
\be \mathcal{F}(x)= \int_0^x f(w) \: \dd w=
\left\{ \begin{array}{lr}
    100-\e^{\displaystyle \e^{(A-Bx)}},
    & (0 \le x<x_t), \\ \\
    100- {(x_t)}^{\alpha} \: \e^{\displaystyle \e^{(A-Bx_t)}}
    x^{\, -\alpha},
    & \; \; \; (x_t\le x<\infty),
    \end{array}
    \right.
    \lb{lorenz-totalX}
\ee
and
\be \mathcal{F}_1(x)=
\frac{1}{\langle x \rangle} \int_0^x w \: f(w) \: \dd w=
\left\{ \begin{array}{lr}
    \displaystyle \frac{\mathcal{I}(x)}{ \langle x \rangle}
    \,, & (0<x<x_t), \vspace{4mm} \\
    100+\displaystyle \frac{\displaystyle \alpha \, {(x_t)}^{\alpha}
    }{\displaystyle (1-\alpha)}
    \: \frac{\e^{\displaystyle \e^{(A-Bx_t)}}}{\langle x \rangle}
    \; {x^{(1-\alpha)}}, & \; \; \; (x_t\le x<\infty;).
    \end{array}
    \right.
    \lb{lorenz-totalY}
\ee
Thus, $\mathcal{F}_1(x)$ varies from 0 to 100 as well. The Lorenz
curve is usually represented in a unit square, but the normalization
(\ref{norm1}) implies that the square where the Lorenz curve is
located has area equal to $10^4$.

The \textit{Gini coefficient} under the currently adopted
normalization is written as,
\be Gini = 1- 2\times 10^{-4} \int_0^\infty \mathcal{F}_1(x)
         \: f(x) \: \dd x.
    \lb{gini}
\ee
Considering now equations (\ref{distro2}) and
(\ref{lorenz-totalY}), the Gini coefficient has the
following expression in the GPD,
\be Gini=1-2\times 10^{-4} \left\{ \frac{B}{\langle x \rangle}
         \int\limits_0^{x_t} \mathcal{I}(x)\: \e^{(A-Bx)}
	 \e^{ \displaystyle \e^{(A-Bx)} } \dd x +
	 100\:
	 \e^{ \displaystyle \e^{(A-Bx_t)} } 
	 +\frac{\alpha^2 \: {x_t} \:
	 \e^{  2 \, \displaystyle \e^{(A-Bx_t)}  } 
	 }{\langle x
	 \rangle (\alpha-1)(1-2\alpha)} \right\}.
	 \lb{gini2}
\ee

As discussed in \cite{fnm10}, we can define the \textit{percentage
share of the Gompertzian part} of an income distribution described
by the GPD by means of equation (\ref{lorenz-totalY}). This quantity
may then be written as follows,
\be u=\mathcal{F}_1(x_t)=100-\frac{\alpha}{(\alpha-1)}
    \frac{x_t}{\langle x \rangle}
       \e^{ \displaystyle \e^{(A-Bx_t)}}.
       \lb{u2}
\ee
Hence, we identity the percentage share of the lower income strata
described by the GPD with Goodwin's labor share $u$. Note that by
doing so, $u$ no longer represents the industrial reserve army of
labor, but in fact the \textit{relative surplus population} since
the latter includes not only the unemployed, but also those unable
to work. Such identification allows the description of the Goodwin
variables in terms of measurable quantities connected to different
income classes whose empirical values can be obtained, for instance,
from the Lorenz curves. This connection can be made clearer by the
inversion of equation (\ref{u2}), 
\be \frac{1}{\alpha}= 1- \left[ \frac{\displaystyle u \,  x_t \,
     \e^{ \displaystyle \e^{(A-Bx_t)} }}{ \displaystyle (100-u)
     \mathcal{I} \, (x_t)} \right] .
	   \lb{alfa}
\ee
Due to the high non-linearity of this expression one can only
use it to determine $\alpha$ if the values of $u$, $B$ and
$x_t$ are known to a very high degree of accuracy.

The equation (\ref{alfa}) links the Pareto index $\alpha$ to parameters
which are solely determined in the Gompertzian segment of the
distribution: the cutoff value $x_t$, the Gompertzian percentage
share $u$ and its distribution slope $B$. In other words, equation
(\ref{alfa}) links the income distribution of the lower and upper
classes forming a society, showing clearly their dynamical
inter-dependency. If we consider that temporal changes in the
income distribution do take place, we can no longer consider these
quantities as parameters. Some of them, or perhaps all of them,
ought to be time dependent variables (see below).

The GPD requires $\alpha>0$. In addition, an average income is
only possible if $\alpha>1$. Considering these two conditions
in equation (\ref{alfa}) we conclude that,
\be 0<\left[ \frac{\displaystyle u \,  x_t \,
     \e^{ \displaystyle \e^{(A-Bx_t)} }}{ \displaystyle (100-u)
     \mathcal{I} \, (x_t)} \right]<1 \; \; \; \; \; \mbox{and}
     \; \; \; \; \; u<100.
\lb{cond}
\ee
Remembering equation (\ref{inv}) the last condition is equivalent
to $U>0$, which means that an income distribution described by
the GPD is only possible in a system where investors have a
nonzero share of the total income.

\subsection{Exponential approximation}\lb{exp}

As shown in Refs.\ \cite{nm09,fnm10}, the upper part of the
Gompertz curve can be approximated by an exponential and this
allows us to take this subdivision of the Gompertz curve as
representing the middle class present in most societies.
In other words, in this approach of the income distribution
characterization of a society we assume that the middle class
is just the upper echelon of the wage labor class. Thus, for
$Bx>A$, $\e^{-Bx}<1$ and $x<x_t$ we have that,
\be
\left\{
    \begin{array}{ll}
    G(x)  \approx 99+  \e^{-Bx}, \\ 
    \mathcal{G}(x)  \approx 1-  \e^{-Bx}, \\ 
    g(x) \,  \approx B \; \e^{-Bx},
    \end{array}
\right.
\lb{gex}
\ee
which are already normalized to obey the boundary conditions
(\ref{condi}).
If the lower stratum of a society is formed essentially by a very
large middle class, one can in principle write all equations shown
in Sect.\ \ref{gpdis} in terms of the approximations (\ref{gex}),
although in such a case we can expect a certain degree of distortion
in the distribution since all modern societies seem to have a certain
percentage of very poor people, however small this percentage may be.

\section{Cycles in the income and employment data of Brazil}
\lb{data-orbits}

Publicly available individual income distribution data of the
Brazilian population have allowed Moura Jr.\ and Ribeiro \cite{nm09}
to determine the GPD parameters from 1978 to 2005 after a careful
handling of the data. Chami Figueira, Moura Jr.\ and Ribeiro
\cite{fnm10} extended this analysis to include income data for 2006
and 2007, as well as showing how the GPD produces results compatible
with those obtained directly from the raw data, that is, without
assuming the GPD, with error margins up to 7\%. In this work we
further extend these two previous analyzes to include data for 2008
and 2009, but disregarding the results for 1978 and 1979 due to their
unreliability \cite{fnm10}.

Table \ref{tab1} presents the three GPD parameters $B$, $x_t$
and $\alpha$ followed by the unemployment rate $[V]$, Gini
coefficient and the percentage share of the Gompertzian component of
the distribution. $B$ and $\alpha$ were obtained by linear data
fitting whereas $x_t$ was determined such that a linear fit would
produce the boundary condition (\ref{a}) with discrepancy of about
2\%. Lorenz curves were generated from the raw distribution for each
year allowing the calculation of the Gini coefficient without
assuming the GPD, denoted here as $[Gini]$ in order to distinguish
it from the one obtained assuming the GPD in equation (\ref{gini2}).
Once $x_t$ was found it became possible determine $[u]$ directly
from the raw data, that is, without using equation (\ref{u2}). Similarly,
$[V]$ denotes the unemployment data without any distribution
assumptions, $[v]$ is obtained using equation (\ref{V}) and $x_d$ is the
\textit{unemployment income threshold} used to calculate $[V]$ (see
below). The time derivatives are given by the expressions,
\be
\dot{[u]}=\frac{\dd}{\dd t} [u] \, , \; \; \; \; \;
\dot{[v]}= \frac{\dd}{\dd t} \left(100 - [V]\right).
\lb{uvponto}
\ee
\begin{table*}[htbp]
\caption{Data for Brazil from 1981 to 2009. Values in brackets mean
         that they were evaluated without using the GPD parameters, that
         is, directly from the raw data. This table contains the GPD
         parameters $B$, $x_t$ and $\alpha$ for the individual income,
         unemployment income threshold $x_d$, unemployment rate $[V]$, Gini
         coefficient, percentage share $[u]$ of the Gompertzian component
         (workers' share), employment rate $[v]$ and time derivatives of
         the last two quantities, as given by equations (\ref{uvponto}). The
         results from 1981 to 2007 had already appeared in \cite{nm09,fnm10}
         whereas those for 2008 and 2009, as well as the ones for employment
         and unemployment, are new. The time derivatives $\dot{[u]}$ and
         $\dot{[v]}$ were calculated numerically using equation (\ref{dfnum}).
         Since there were no income samplings in 1991, 1994 and 2000
         \cite[see][]{nm09}, some results for these years were obtained by
         numerical interpolation.\label{tab1}}
\begin{center}
\small
\begin{tabular}{ccccccccccc}
\hline\noalign{\smallskip}
year & $B$ & $x_{\sty t}$ & $\alpha$ &$x_d$&$[V] \, (\%)$ & $[Gini]$ &
$[u] \, (\%)$ & $\dot{[u]} \, (\%/\mbox{year})$ & $[v] \, (\%)$ &
$\dot{[v]} \, (\%/\mbox{year})$ \\ 
\noalign{\smallskip}\hline\noalign{\smallskip}
1981 &$0.342\pm0.016$&7.533&$2.839\pm0.091$ &$0.182$& $14.8$ & $0.574$ &
$87.7$ & &$85.2$&       \\
1982 &$0.342\pm0.015$&7.473&$2.677\pm0.042$ &$0.174$& $14.5$ & $0.581$ &
$87.2$ & $+1.08$ &$85.5$& $-0.20$ \\
1983 &$0.330\pm0.010$&6.910&$2.636\pm0.081$ &$0.175$& $14.5$ & $0.584$ &
$85.5$ & $-0.04$ &$85.5$& $-1.06$ \\
1984 &$0.332\pm0.013$&7.388&$2.839\pm0.072$ &$0.170$& $12.4$ & $0.576$ &
$87.2$ & $-0.17$ &$87.6$& $-1.32$ \\
1985 &$0.329\pm0.010$&7.490&$2.656\pm0.093$ &$0.154$& $11.8$ & $0.589$ &
$85.8$ & $+0.99$ &$88.2$& $-2.22$ \\
1986 &$0.344\pm0.013$&7.112&$2.567\pm0.065$ &$0.127$& $ 7.9$ & $0.580$ &
$85.2$ & $-0.05$ &$92.1$& $-1.16$ \\
1987 &$0.343\pm0.016$&7.626&$2.724\pm0.057$ &$0.127$& $ 9.5$ & $0.592$ &
$85.9$ & $-0.08$ &$90.5$& $+2.11$ \\
1988 &$0.324\pm0.014$&8.140&$2.874\pm0.125$ &$0.133$& $12.1$ & $0.609$ &
$85.4$ & $+1,74$ &$87.9$& $+0,17$ \\
1989 &$0.317\pm0.010$&7.856&$2.428\pm0.079$ &$0.111$& $ 9.9$ & $0.628$ &
$82.5$ & $-0.23$ &$90.1$& $-2.35$ \\
1990 &$0.335\pm0.015$&8.074&$2.636\pm0.053$ &$0.099$& $ 7.4$ & $0.605$ &
$85.9$ & $-1.98$ &$92.6$& $+0.48$ \\
1991 &               &     &                &       & $10.8$ &         &
$86.4$ & $-0.57$ &$89.2$& $+3.37$ \\
1992 &$0.364\pm0.020$&7.635&$2.636\pm0.063$ &$0.162$& $14.2$ & $0.578$ &
$87.0$ & $+1.18$ &$85.8$& $+0.53$ \\
1993 &$0.330\pm0.008$&7.674&$2.567\pm0.042$ &$0.137$& $11.9$ & $0.599$ &
$84.1$ & $+1.01$ &$88.1$& $-2.54$ \\
1994 &               &     &                &       & $ 9.1$ &         &
$85.0$ & $-0.92$ &$90.9$& $-2.75$ \\
1995 &$0.333\pm0.012$&7.887&$2.777\pm0.106$ &$0.098$& $ 6.4$ & $0.596$ &
$85.9$ & $-0.86$ &$93.6$& $-0.63$ \\
1996 &$0.347\pm0.020$&8.163&$2.749\pm0.107$ &$0.096$& $ 7.8$ & $0.598$ &
$86.7$ & $-0.12$ &$92.2$& $+0.43$ \\
1997 &$0.338\pm0.015$&7.935&$2.617\pm0.052$ &$0.099$& $ 7.2$ & $0.598$ &
$86.1$ & $+1.09$ &$92.8$& $-0.29$ \\
1998 &$0.326\pm0.009$&7.628&$2.677\pm0.031$ &$0.103$& $ 7.3$ & $0.597$ &
$84.5$ & $+0.08$ &$92.7$& $+0.25$ \\
1999 &$0.331\pm0.013$&7.811&$2.777\pm0.068$ &$0.107$& $ 7.7$ & $0.590$ &
$86.0$ & $-0.53$ &$92.3$& $+0.55$ \\
2000 &               &     &                &       & $ 8.4$ &         &
$85.6$ & $+0.40$ &$91.6$& $+0.66$ \\
2001 &$0.335\pm0.011$&7.774&$2.724\pm0.205$ &$0.122$& $ 9.0$ & $0.592$ &
$85.2$ & $-0.41$ &$91.0$& $-0.24$ \\
2002 &$0.339\pm0.015$&7.878&$2.500\pm0.121$ &$0.123$& $ 7.9$ & $0.586$ &
$86.4$ & $-0.08$ &$92.1$& $+0.01$ \\
2003 &$0.333\pm0.009$&7.374&$2.777\pm0.057$ &$0.134$& $ 9.0$ & $0.579$ &
$85.4$ & $-0.40$ &$91.0$& $-1.07$ \\
2004 &$0.333\pm0.017$&8.005&$3.234\pm0.133$ &$0.105$& $ 5.7$ & $0.582$ &
$87.2$ & $-0.44$ &$94.3$& $-0.59$ \\
2005 &$0.326\pm0.009$&7.403&$2.839\pm0.089$ &$0.118$& $ 7.9$ & $0.580$ &
$86.2$ & $-0.26$ &$92.1$& $+2.06$ \\
2006 &$0.323\pm0.015$&8.078&$3.749\pm0.136$ &$0.125$& $ 9.9$ & $0.592$ &
$87.7$ & $+0.27$ &$90.1$& $-0.29$ \\
2007 &$0.334\pm0.009$&6.934&$2.839\pm0.104$ &$0.125$& $ 7.3$ & $0.572$ &
$85.7$ & $+0.28$ &$92.7$& $-1.03$ \\
2008 &$0.366\pm0.011$&6.848&$2.567\pm0.051$ &$0.141$& $ 7.8$ & $0.543$ &
$87.2$ & $-0.36$ &$92.2$& $+0.26$ \\
2009 &$0.363\pm0.010$&6.500&$2.656\pm0.065$ &$0.148$& $ 7.8$ & $0.539$ &
$86.4$ & &$92.2$&       \\
\noalign{\smallskip}\hline
\end{tabular}
\end{center}
\end{table*}

One should note that the focus of this paper is not to discuss the
adequacy of the GPD description of income data by comparing results
obtained by assuming or not the GPD, that is, comparing $Gini$ to
$[Gini]$ or $u$ to $[u]$, as this task was already successfully
accomplished in Ref.\ \cite{fnm10}. Our focus here is to use the
GPD as a tool to partition the income distribution in the
Gompertzian and Paretian components, identify the former with one
of the variables of the Goodwin model and to discuss the possible
dynamical implications of such a division, that is, linking the
GPD parameters to the Goodwin model. The unemployment data
appearing in Table \ref{tab1} require, however, some explanation
about how they were determined.

Two basic facts prevented us from using official Brazilian joblessness
statistics in the analysis studied here. First,
unemployment data collection methodology changed quite substantially
during the time period of this study (1981 to 2009) and, secondly,
its sampling method differs from the one used to survey income. Taken
together, these two facts imply the lack of sample homogeneity in the
whole period of this analysis, which renders it impossible to derive
measurable quantities without introducing substantial statistical
biases. Without sample homogeneity we cannot compare unemployment data
from early and late years in the studied time interval. These
difficulties can be avoided if unemployment is directly estimated from
the income database by means of a criterion applicable to the entire
time period of this study. The reasoning we followed to do that is
described below.

Every society produces useful energy and materials to be consumed by
the people who participate in their production. This means that a
person active in this production receives a share of those materials
and energy, that is, a share of the total value produced by the
society in a certain period of time. Income is, therefore, a flow of
value (energy and materials) a person receives in a certain time
period. Under this viewpoint, even food is part of this share. The
unemployed is the individual who does not participate in the
production and, therefore, does not receive value. Nevertheless,
nobody can survive too long without food or a minimum amount of
energy and, thus, if the individual survives this means that somehow
this individual still has a value inflow. Such a minimum supporting
value is usually provided by family or, in more limited ways, by the
state, but actually means a reduced value inflow for the group family
this individual belongs. In other words, when somebody becomes
unemployed those close to this individual are the ones who suffer most
because the whole family has a smaller share of value or, which is
stating the same, the group family income decreases. So, there should
be a limit in income distribution where unemployment, or underemployment,
can be effectively detectable. We call this limit \textit{effective
unemployment}. An \textit{average} person who receives up to this
minimum income barely participates in the production and for all
practical effects is jobless.

Following this reasoning we then probed the data for income values
which would produce unemployment rates \textit{in agreement} to those
in the official unemployment surveys for the last 15 years or so. Our
results showed that effective unemployment occurs when the average
individual income is equal to or below 20\% of the national minimum
salary in Brazil expressed in US dollars and effective at the time the
income survey was carried out (September of each year). This amount
defines the \textit{unemployment income threshold} $x_d$ which, after
being normalized to become a currency free quantity, was
applied to the income distribution of each year to obtain the
percentage share of those in the distribution whose income were equal
to or below this amount. This method provides our effective definition
of unemployment.

Connecting the unemployment income threshold $x_d$ to minimum salary
has the advantage of providing a simple criterion applicable to income
data for all years of this study, even before 1994 when Brazil sampled
unemployment through a different methodology and experienced runaway
inflation and hyperinflation. The results for the unemployment rate
$[V]$ obtained using this criterion are presented in Table \ref{tab1}.
Note that once $x_d$ is known, the GPD allows us to obtain $V$ by means
of an expression similar to equation (\ref{u2}). Indeed, as one should have
$x_d < x_t$, remembering equation (\ref{lorenz-totalY}) we conclude that,
\be
 V=\mathcal{F}_1(x_d)=\frac{I(x_d)}{\langle x \rangle}.
\lb{v2}
\ee

We can now plot the results for $[u]$ and $[v]$. Figure \ref{uvtime}
shows the time evolution of these two variables where one can see 
that both variables cycle with similar periods of about 4 years. In
addition, these cycles are apparently out-of-phase for most of the
studied time interval and have phase difference of about 2 years. This
clearly implies short term cycles. These results bring qualitative
empirical support to the Goodwin approach for describing the dynamics
of the capitalist production as described by Marx.

Figure \ref{uv-total} shows the $[u]$-$[v]$ phase space where one can
see clockwise orbits for most of the time interval, a fact which again
brings qualitative empirical support to the Goodwin model at least as
far as Brazilian data is concerned. However, the orbits clearly do not
have a single center, as both the original Goodwin model and its DHMP
extension predict. After 1994 the center of the orbits seems to move
to an upper position in the phase space. In order to better appreciate
this change, figure \ref{uv-parts} shows the same results of figure
\ref{uv-total}, but divided in two time intervals, from 1981 to 1994 and
1995 to 2009. These results clearly contradict the Goodwin prediction
of all orbital centers having the same fixed coordinates $u_c$ and $v_c$,
as described by equation (\ref{center}). One should also note that these
results are entirely different from the ones obtained for Brazil in
Ref.\ \cite{mome} using Harvie's method \cite{harvie} and in much
better agreement at a qualitative level with the Goodwin model. Finally,
figure \ref{3d} shows the same data in a 3-dimensional plot with the
Z-axis representing the time. This graph provides a different way of
seeing the displacement of the points to a different region after 1994
by means of their projection in the YZ-plane, as well as a possible
earlier displacement, whose transition occurred from 1981 to 1983.
\begin{figure}[ht]
\begin{center}
\includegraphics[scale=1.0]{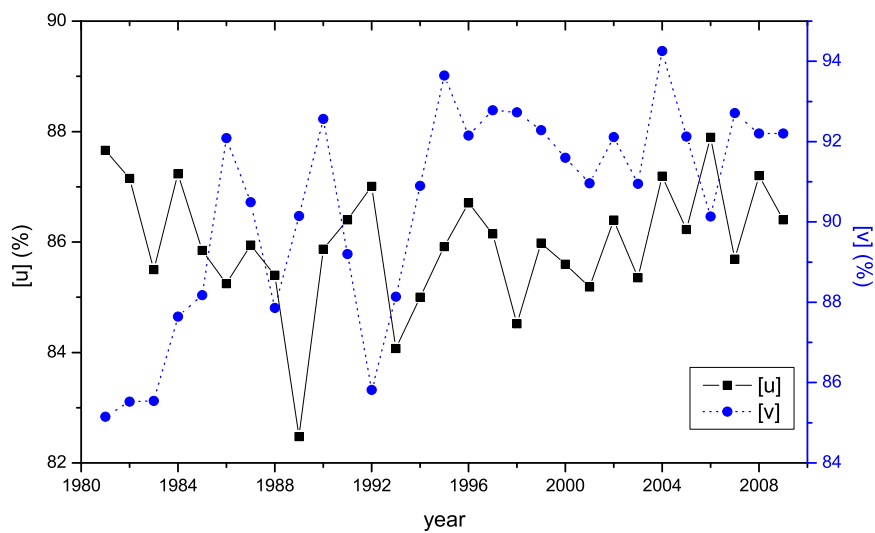}
\caption{Time evolution of the Gompertzian component (workers' share)
         $[u]$ and employment rate $[v]$ in Brazil. The plot shows that
         these variables cycle out-of-phase for most of the studied 
         time interval with periods of about 4 years in both variables,
         meaning that booms and busts in Brazil occur in short term
         cycles. These results show that predator-prey like models can
         be used to represent real economic systems.}\lb{uvtime}
\end{center}
\end{figure}

\begin{figure}[ht]
\begin{center}
\includegraphics[scale=1.2]{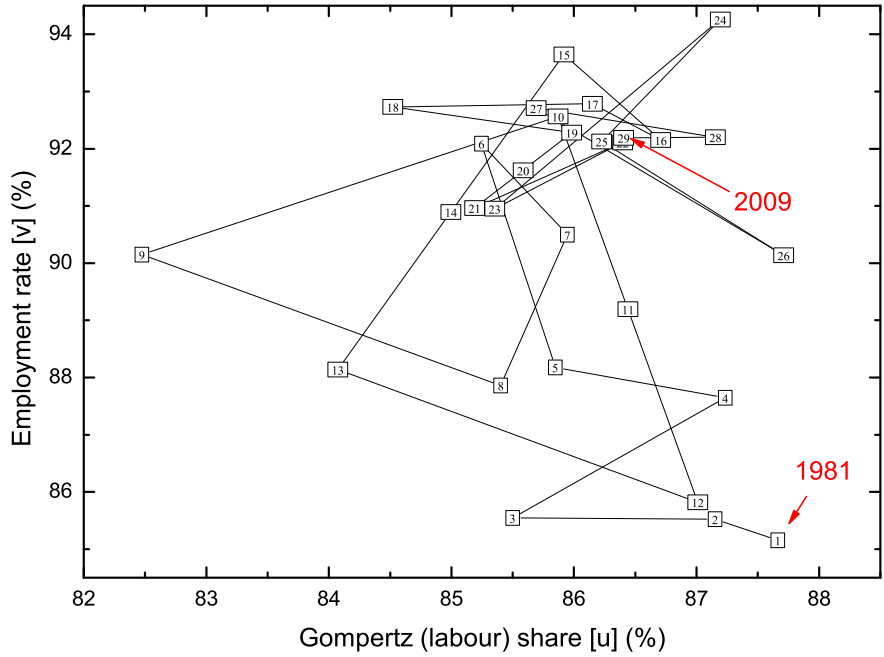}
\caption{$[u]$-$[v]$ phase space for Brazil from 1981 to 2009. The plot
         points are labeled in growing numerical sequence with each
         integer number representing one year in the studied time
         interval. Thus, the label given by the number `1' indicates
         the year 1981 and label `29' means the year 2009, providing
         then a clear visualization of the largely clockwise evolution
         of the cycles. One can see that before 1995, indicated by label
         `15', the system was cycling in a different region of the
         phase space. The end of hyperinflation in 1994 (label `14') is
         possibly the event which made the system move to a new cycling
         region, where it still remains.}\lb{uv-total}
\end{center}
\end{figure}

\begin{figure}[ht]
\begin{center}$
\begin{array}{cc}
\includegraphics[scale=0.83]{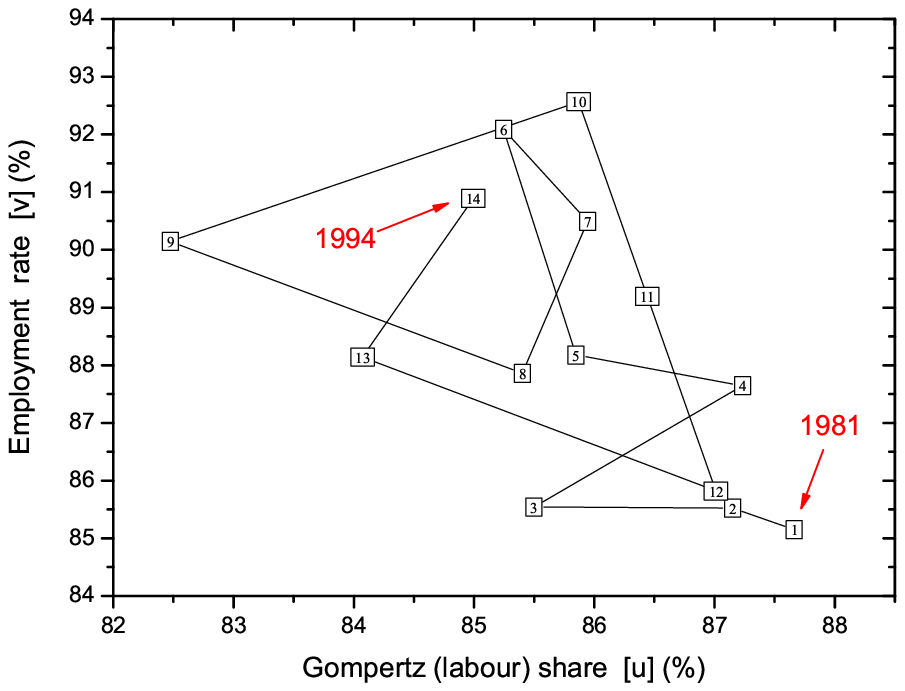} &
\includegraphics[scale=0.70]{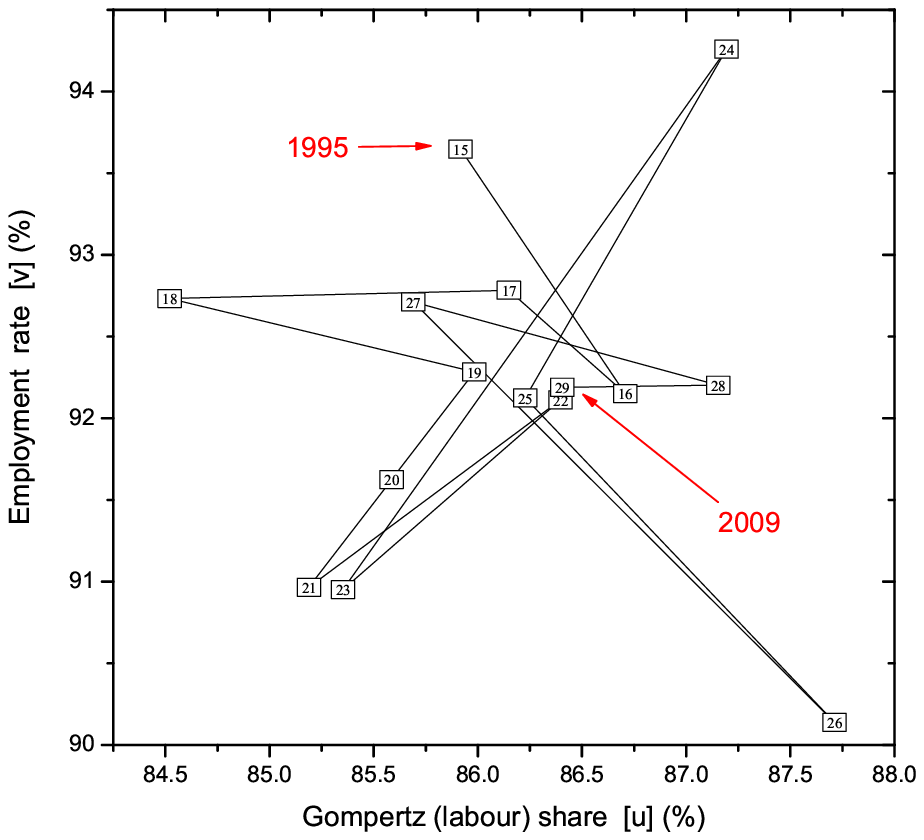} 
\end{array}$
\end{center}
\caption{These two graphs present the same data as in Fig.\ \ref{uv-total},
         but divided into two sets of points. The plot in the \textit{left}
         shows the $[u]$-$[v]$ phase space from 1981 until 1994, whereas
         the \textit{right} plot presents the data points from 1995 to
         2009. This shows even more clearly that the Brazilian economic
         system moved from one region to another in the phase space during
         the time interval studied here. One can also note in the left plot
         that the system was possibly moving from yet another region in
         the period from 1981 to 1983, since the labels `1' to `3'
         indicating these years appear to be part of a transition from a
         different region than the one where the system remained until
         1994. This possible interpretation has some empirical support
         because the high inflationary period in Brazil started in about
         1980.}
\lb{uv-parts}
\end{figure}

\begin{figure}[ht]
\begin{center}
\includegraphics[scale=1.2]{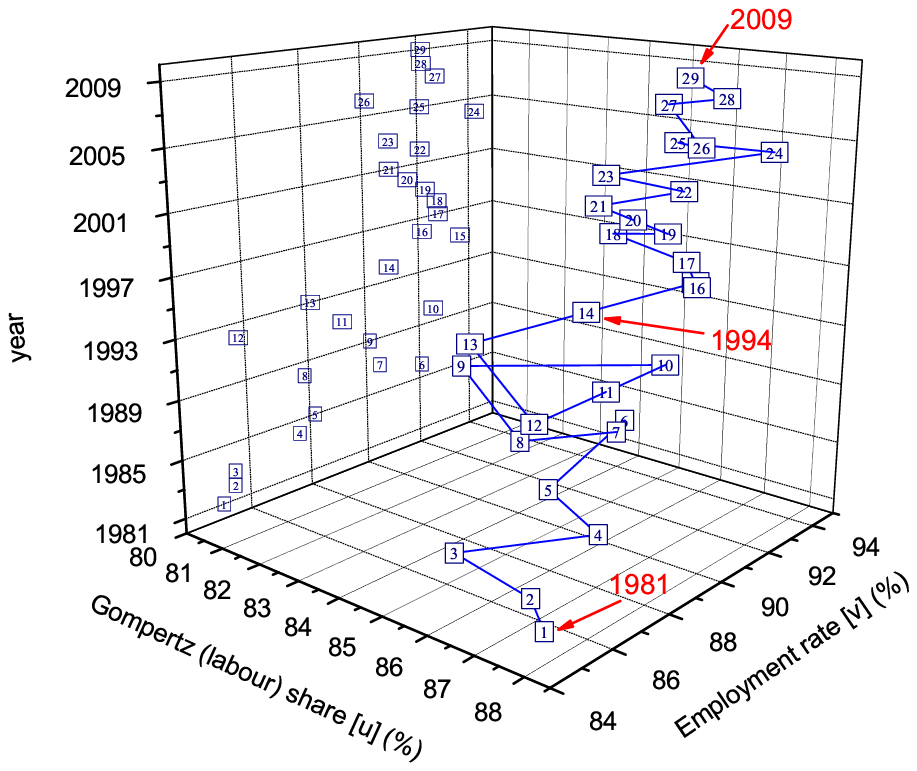}
\caption{This plot is a 3-dimensional representation of the same
         points appearing in figures \ref{uv-total} and \ref{uv-parts}.
         It provides a different visualization of the system displacements
         during its evolution from 1981 to 2009, showing more clearly
         in the YZ-plane projection (the surface plane for $[v]$ vs.\
         time, on the left side of the plot) the three regions where
         the system located itself in the studied time interval. The
         points for 1981 to 1983 (labels `1' to `3') seem to be a
         transition from an unspecified earlier region where the system
         stayed before the high inflationary period started at about
         1980. The end of hyperinflation in 1994 (label `14') moved the
         system to yet another region on the top right of the YZ-plane.}
\lb{3d}
\end{center}
\end{figure}

The important event which may explain why the apparent orbital center
changes location after 1995 is the end of hyperinflation. In 1994 Brazil
established a new and stable currency, the real (R\$), which abruptly
ended the strong inflationary period of the previous 15 years. This fact
seems to be reflected in the $[u]$-$[v]$ phase space by a change in the
center of the orbit. One can also see in table \ref{tab1} a slow,
although modest, decrease in the Gini coefficient after 1993. In addition,
since the Brazilian high inflationary period started at about 1980, the
positions corresponding to the years of 1981 to 1983 in the phase space
appear to represent an earlier transition from yet another region in
the phase space. This seems to be the case if we carefully look at these
points in the graphs of figures \ref{uv-total} and \ref{3d}.

The absence of a single center for all orbits means that the parameters
$a$, $b$, $c$, $d$, and $h$ of the Goodwin model are most likely not
constants at all, but time dependent variables. Nevertheless, at a
qualitative level the model certainly has empirical support which
justifies the identification of $[u]$ and $[v]$ with $u$ and $v$,
although in order to understand the real dynamics behind these
quantities one probably needs to somehow modify the dynamical equations
(\ref{u1}) and (\ref{v1}) to reflect these empirical evidences.

Finally, we should note that the lack of a single orbital center in
real-world data has already appeared in earlier empirical studies
carried out by other authors on the Goodwin model. The $u$-$v$
phase-space plots of Desai \cite{desai}, Solow \cite{solow}, Harvie
\cite{harvie}, Moreno \cite{moreno}, Mohun \& Veneziani \cite{simon}
and Garc\'{\i}a Molina \& Herrera Medina \cite{mome} show similar
results as ours. Vadasz \cite{va} also reached a similar conclusion,
although by indirect means. The important point is that all these
authors reached the same conclusion despite their use of very
different methods to analyze observational data. Therefore, one feels
justified to conclude that this feature appears to be universal and
clearly indicate that the Goodwin model must be changed in order to
accommodate this real-world feature.

\section{Temporal variation of the employment rate and workers'
         share}\lb{5}

The data presented in table \ref{tab1} allow us to go beyond the
qualitative discussion of the previous section and carry out a
quantitative evaluation of the Goodwin model and its DHMP extension.
To do so we need first to carry out simple numerical estimations of
the time derivative $\dot{[u]}$. This task is most straightforwardly
accomplished using the following expression,
\be
\dot{[u]} \approx \frac{[u](t + \Delta t) - [u](t - \Delta t)}{2\Delta t},
\lb{dfnum}
\ee
where $\Delta t =1$ year. Similar procedure is used to determine
$\dot{[v]}$. The goal here is to use data fitting to estimate the
parameters of the two sets of dynamical equations, the first set being
given by equations (\ref{lv1}) and (\ref{lv2}) of the original Goodwin model
and the second one by equations (\ref{lv-d1}) and (\ref{lv-d2}) which
constitute the DHMP extension.

\subsection{Goodwin model}

Figure \ref{original} shows two plots, the \textit{left} one for the
variables of equation (\ref{lv1}) and the \textit{right} plot for equation
(\ref{lv2}). The fitted straight lines parameter values are also
presented for both plots. It is clear that both sets of points are
compatible with a linear approximation similar to the original
dynamical equations, but the parameters behave in exactly opposite
manner from what the model predicts. While the slope of the lines
predicted by equations (\ref{lv1}) and (\ref{lv2}) are, respectively,
positive and negative, the results coming from Brazilian real-world
data are the other way round. This is clear in both graphs. This
result can also be seen if we use the fitted parameters to obtain 
conditions which the supposedly ``constants'' of the Goodwin model
should obey. Doing so we conclude that the Brazilian economic dynamics
studied here gives,
\begin{equation}
 \left\{ \begin{array}{ll}
              c<0, \\
	      h<0, \\
	      (a+d)<0, \\
	      (a+b) \, c>100.
         \end{array}
  \right.
\label{pars3}
\end{equation}
These results completely upset the parameter conditions given by equations
(\ref{pars}), which were thought to be valid. The fitting also leaves
two parameters yet to be determined by some yet unknown equation relating
them since, as seen above, the orbital center and period equations
(\ref{center}) are clearly invalid in the Brazilian income dynamics.
\begin{figure}[ht]
\begin{center}$
\begin{array}{cc}
\includegraphics[scale=0.75]{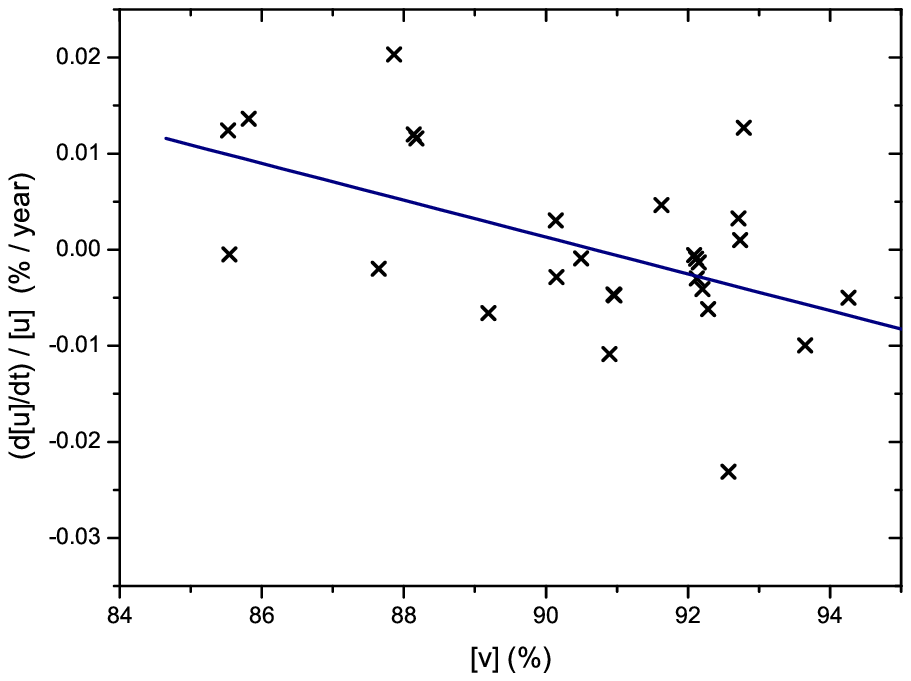} &
\includegraphics[scale=0.75]{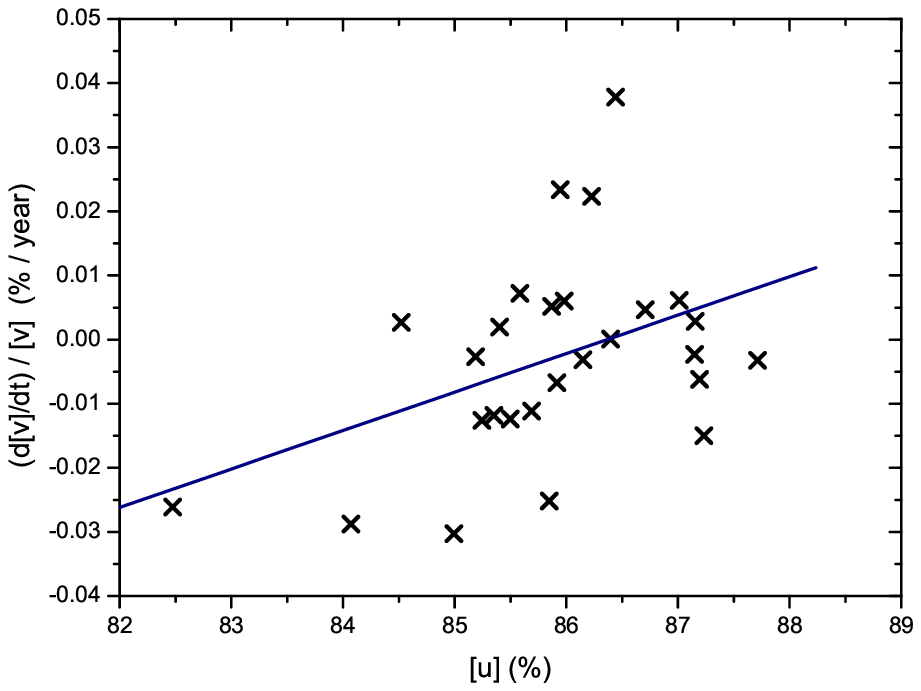} 
\end{array}$
\end{center}
\caption{\textit{Left}: $\dot{[u]}/[u]$ vs.\ $[v]$. \textit{Right}:
    $\dot{[v]}/[v]$ vs.\ $[u]$. Although both graphs show some
    dispersion of the results, one can clearly identify a general
    tendency for the observational points to decrease in the left
    plot and to increase in the right one. Straight line fits on
    both sets of results, indicated as full lines, produced the
    following results. For the \textit{left} plot, the expression
    $\dot{[u]}/[u]=A_1+B_1[v]$ resulted in $A_1=0.17\pm0.06$,
    $B_1=-0.0019\pm0.0006$. For the \textit{right} plot, the
    equation $\dot{[v]}/[v]=A_2+B_2[u]$ yielded parameters as
    follows: $A_2=-0.52\pm0.22$, $B_2=0.006\pm0.003$. These results
    should be compared with equations (\ref{lv1}) and (\ref{lv2}).}
\lb{original}
\end{figure}

The calculated uncertainties in the fitted parameters do not change
this situation, a fact which forces us to conclude that the
economic hypotheses advanced by Goodwin to derive his model are either 
not applicable, partially or completely, to the economic system
studied here or they are flawed. Whatever conclusion one may
choose, this analysis indicates that to advance this model with the
aim of turning it into a viable representation of the real world, the
focus must lie on the probable modification of the set of differential
equations (\ref{lv1}) and (\ref{lv2}) and their empirical validation,
rather than how they were obtained. Only after a good model is achieved,
and by good we mean a model with solid empirical foundations, may we
start looking for the real economic conditions behind its dynamics.

Since the data show that the parameters of the model follow the exact
opposite predictions given by the expressions (\ref{pars}),
another consequence of the results shown in figure \ref{original} is
the reversal of the predicted roles of predator and prey discussed in
Sect.\ \ref{growth}. Indeed, according to the fitted parameters (see 
caption of figure~\ref{original}), when $[u]=0$, $\dot{[u]}=0$ and
$\dot{[v]}<0$. In this case $[u]$ plays the role of prey because without
it ($[u]=0$) the predator population decreases ($\dot{[v]}<0$). Similarly,
when $v=0$, $\dot{[v]}=0$ and $\dot{[u]}>0$. So, $v$ plays the role of
predator because without them ($v=0$) the prey population grows without
bounds ($\dot{[u]}>0$). Such a reversal of roles of predators and preys
coming from the real-world data analysis presented here also implies a
reversal of the reasoning presented in Sect.\ \ref{conflict} regarding
how one interprets the conflicting variables. However, although such
role discussions had some importance in the past, such interpretations
are now of lesser importance than revealing the inner dynamics of these
two inter-dependent variables. When such dynamics is better understood
by means of realistic, not introspective, models, such roles will
naturally emerge from those real-world representations. 

\subsection{DHMP extension}

The variables in the dynamical equations (\ref{lv-d1}) and (\ref{lv-d2})
of the DHMP extension are plotted in figure \ref{figdhmp}. To do so we
had to choose a value for the maximum share of labor $\bar{[u]}$. From
table \ref{tab1} we see that the highest value in the studied time
period is 87.7\% in 1981 and in view of the fact that the DHMP model
does not give any hint about how to obtain $\bar{[u]}$, assuming whatever
constant value higher than that is enough for our purposes here and will
not change the general behavior of equation (\ref{lv-d2}). So, we chose
$\bar{[u]}=95\%$ as a reasonable value for this analysis.

The \textit{left} plot in figure \ref{figdhmp} shows the points related
to the dynamical variables of equation (\ref{lv-d1}) while the
\textit{right} graph in concerned with the variables appearing in
equation (\ref{lv-d2}). The fitted parameters are written in the figure
caption and, similarly to our reasoning above, they produce real-world
conditions for the ``constants'' of the DHMP model. They may be written
as follows,
\begin{equation}
 \left\{ \begin{array}{ll}
          \delta>0 \; \; (?), \\
          \bar{h}>0 \; \; (?), \\
          (\bar{a}+\bar{d})>0 \; \; (?), \\
          \lambda<0, \\
          (\bar{a}+\bar{b})< -\lambda \ln (100- \bar{u}). \\
         \end{array}
  \right.
\label{pars4}
\end{equation}
The results with a question mark are inconclusive due to the uncertainties
of the fitted parameters. For the other two, $\lambda<0$ contradicts the
prediction given in equations (\ref{pars2}), but implies that
$(\bar{a} +\bar{b})>0$ for the chosen $\bar{u}$. So, despite the fitting,
the DHMP model remains in a very inconclusive status regarding the 
empirical behavior of its dynamical variables and its supposedly constant
parameters. Even so, because the model has too many parameters, after a
successful fitting where one of the parameters had to be assumed
($\bar{u}$), two other parameters remained unknown and still
require determination by at least another, also unknown, expression. 

In conclusion, because the DHMP extension has more unknown quantities
and its dynamics is described by somewhat more complex differential
equations than the original Goodwin model, comparing its predictions
with the Brazilian data renders mixed and inconclusive results.
Adding to this situation are the high errors in the fitted
parameters and the fact that even after a successful fit several
parameters remain unknown. These results place the DHMP extension in
a much less favorable situation than the original Goodwin model
regarding empirical validity, at least as far as Brazilian data is
concerned.
\begin{figure}[ht]
\begin{center}$
\begin{array}{cc}
\includegraphics[scale=0.75]{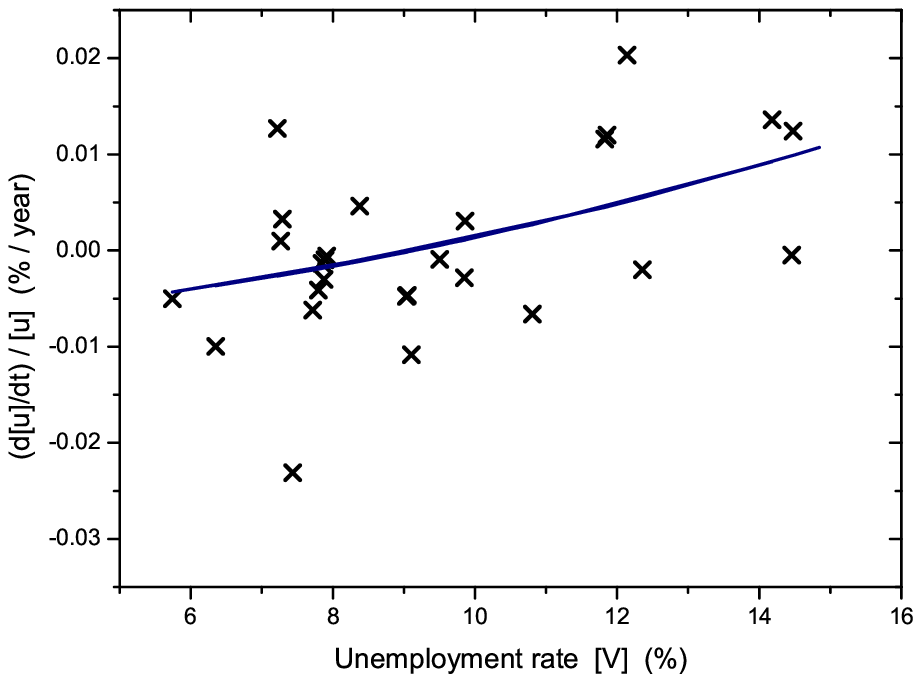} &
\includegraphics[scale=0.75]{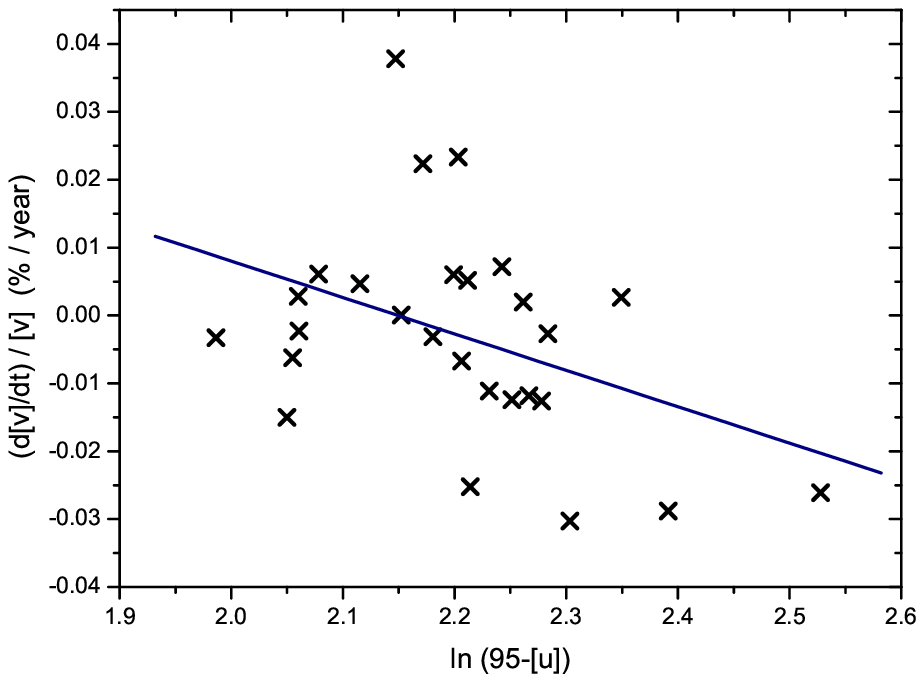} 
\end{array}$
\end{center}
\caption{\textit{Left}: $\dot{[u]}/[u]$ vs.\ $[V]$. \textit{Right}:
    $\dot{[v]}/[v]$ vs.\ $[\mathcal{U}]$ (see equations \ref{U} and
    \ref{V}), where $\bar{u}$ was assumed as 95\%. In the
    \textit{left} plot the full line indicates a power-law fit of
    the form $\dot{[u]}/[u]=\bar{A}_1+\bar{B}_1[V]^{\bar{\delta}}$.
    The fitting parameters yielded $\bar{A}_1=-0.011\pm0.022$,
    $\bar{B}_1=0.0003\pm0.0027$ and $\bar{\delta}=1.59\pm2.94$,
    results which should be compared with equation (\ref{lv-d1}). For
    the \textit{right} graph the full line indicates a straight
    line fit using the expression $\dot{[v]}/[v]=\bar{A}_2
    +\bar{B}_2 \ln (95-[u])$. The resulting fitted parameters are 
    $\bar{A}_2=0.12\pm0.05$ and $\bar{B}_2=-0.054\pm0.024$. This 
    result should be compared with equation (\ref{lv-d2}).}
\lb{figdhmp}
\end{figure}

\section{Conclusions}\lb{fim}

In this paper we have studied the empirical validity of the model
of economic growth with cycles advanced by Goodwin \cite{g67,gan97}
and one of its specific variations, the Desai-Henry-Mosley-Pemberton
(DHMP) extension \cite{dhmp}, using Brazilian income data from 1981
to 2009. The variables used by Goodwin in his model, the workers'
share of total production $u$ and employment rate $v$ were obtained
by describing the individual income distribution by the Gompertz-Pareto
distribution (GPD) \cite{fnm10}, formed by the combination of the
Gompertz curve, representing the overwhelming majority of the
population ($\sim \,$99\%), with the Pareto power law, representing
the tiny richest part ($\sim \,$1\%) \cite{nm09}. We identified the
Gompertzian part of the distribution with the workers and the
Paretian component with the class of capitalists and used GPD
parameters obtained for each year in the studied time period to
analyze the time evolution of these variables by means of the Goodwin
dynamics. Unemployment data was also obtained from income distribution
so that all variables come from the same sample since Brazilian
unemployment data was collected under different methodologies
during the time span analyzed here.

The results were, however, mixed, both qualitatively and
quantitatively. The data showed clockwise cycles in the $u$-$v$
phase space in agreement with the model, but those cycles were only
largely clockwise and the orbital center was not unique, results
which brought only partial qualitative agreement of the model with
Brazilian data. We obtained temporal variations of the variables and
their derivatives and carried out straight line fittings to the points
formed with these quantities, both in the original Goodwin model and
its DHMP extension in order to obtain fitting parameters which were
compared with predictions of both models. In this respect the original
model was able to provide a better empirical consistency, but the
observed parameters were different from what the model predicts in the
sense of their general behavior, leading to fitted lines whose slopes
had opposite behavior than the theory states. A similar situation
occurred with the DHMP extension, but in this case the uncertainties
in the fitted parameters were too large, leading to mostly inconclusive
results. Although a general predator-prey like behavior was observed,
the lack of a single orbital center and parameters behaving very
differently from what was anticipated bring into question the economic
hypotheses used by Goodwin in deriving his model. It appears that they
may be inapplicable to the economic system under study, a conclusion
which comes as no surprise in view of the extremely simple specifications
of the model, as discussed in Sect.\ \ref{origin}. 

Considering these results, in order to provide a viable representation
of the real world the Goodwin model must be modified. Firstly, as it
is obvious from our results, as well as the ones obtained by previous
authors, there cannot be a single orbital center. We can envisage two
possible reasons for such a result: \textit{(i)} the ``constants'' of
the model may not be constants at all, but time variables; \textit{(ii)}
the right-hand side of equations (\ref{lv1}) and (\ref{lv2}) are too
simple and may require more terms involving the two variables, which
means giving up the linear approximation of equations (\ref{o8}) and
(\ref{o13}). In other words, going to a fully nonlinear modeling.

Secondly, the emphasis so far given by several studies on the
economic foundations of the model, which have been the main
source for its proposed theoretical modifications,
should be put aside, at least temporarily, in favor of devising
differential equations capable of reproducing the observed features,
like the moving orbital centers and the behavior of the graphs with
the temporal variations of $u$ and $v$. Clearly those economic
hypotheses will need to be revised as they produce a model which
does not agree with the data, but these revisions must be made in the
light of empirical results and not solely by theoretical introspection.
Possibly new variables representing other economic players, like debt
and government policy, may have to be introduced in the model, which
means that, perhaps, more than two coupled differential equations would
be necessary to define the economic system. In this respect, as discussed
by Keen \cite{keen95}, Hudson \cite{bubble} and Hudson \& Bezemer
\cite{hb}, investment is not profit, being debt-financed when it exceeds
profit, and government taxation has to be deduced from output to determine
profit. 

Thirdly, since the DHMP model fared much more poorly as compared
to the original model, Occam's razor dictates that these modifications
must be focused in the latter rather than the former because the original
model is simpler. So, developing more complex models without a clear
empirical motivation, and in the absence of a clear guidance given 
by real-data observations, goes against Occam's razor.

The basic motivation behind these proposed modifications comes from
the realization that in its present state the Goodwin model does not
provide much more explanatory power beyond the original qualitative ideas
advanced by Marx. This is so because it is essentially a
mathematical dressing of Marxian ideas by means of a predator-prey
set of first order differential equations, but which produces
solutions that clearly contradict empirical data in many respects
and provides only general qualitative agreement with real-world observations.
Therefore, the real challenge lies in devising a model that addresses
real-world data and is capable of surviving empirical verification.
One must always keep in mind that the good scientific practice entails
a permanent search of convergence between hypotheses and evidences.

\vspace{5mm}
\small
Our thanks go to E.\ Screpanti for the initial encouragement to
pursue this research, S.\ Sordi for pointing out relevant
bibliographic information at the beginning of this project and M.\
Desai and A.\ Kirman for discussions. We are also grateful to S.\ Keen
for various very interesting and useful insights on the origins of the
Goodwin model and the referees for useful comments. One of us (MBR)
acknowledges partial financial support from the Rio de Janeiro State
funding agency FAPERJ.

\end{document}